\documentclass[epj-spec]{svjour}

\usepackage{graphicx,amsmath}
\begin{document}
\title{Extending the Reach of Hydrodynamics}
\author{Scott Pratt}
\institute{Department of Physics and Astronomy,
Michigan State University\\
East Lansing, Michigan 48824-1321}
\date{\today}

\abstract{
Recent and ongoing improvements to hydrodynamic treatments at RHIC are extending the physics reach of hydrodynamics, and improving the phenomenology. Here, the links between technological improvements and the extension of physics are emphasized.  
}
\maketitle

Hydrodynamic models represent the foundation for RHIC modeling. Using experimental data to address the equation of state, viscosity or dynamics of the novel matter created at RHIC would be impossible without hydrodynamic treatments. The last few years have seen a renaissance in the technical development of hydrodynamical calculations. Improvements have included: solving fully three-dimensional equations, incorporating viscosity, coupling to dynamic mean fields. In this talk I will both describe each technical improvement and describe the physical implications and phenomenological consequences of the additional functionality.

\section{Solving fully three-dimensional equations}

Most of RHIC phenomenology has been confined to the central unit of rapidity. By assuming Bjorken-Hwa boost invariance, many hydrodynamic calculations have eliminated the need to model motion along the beam ($z$) axis by exploiting the independence of the equations for translations in the variable $\eta$, defined by:
\begin{equation}
z=\tau\sinh(\eta),~~t=\tau\sinh(\eta).
\end{equation}
After hydrodynamic equations are translated from $x,y,z,t$ to $x,y,\eta,\tau$, it is assumed that all quantities are independent of $\eta$ which makes the equations effectively two-dimensional. In the boost-invariant ansatz the rapidity $y$ of the matter is set equal to $\eta$ which corresponds to acceleration-less matter, an assumption only warranted if there is a broad rapidity distribution. Otherwise, pressure gradients along the $z$ direction would alter the evolution, and even affect the matter emitted at mid-rapidity. 

For observables outside central rapidity, the extension to three dimensions is not an improvement -- it is a necessity, and I will not elaborate on the obvious benefits. Within the central unit of rapidity, the significance of performing three-dimensional calculations is subtle. First of all, the rapidity distribution at RHIC is consistent with thermal sources whose velocities are confined with $\sim$ 1.5 units of rapidity of $y=0$. Since a source moving with 1.5 units of rapidity can still emit a few percent of its particles with zero rapidity, the diligent modeler should modulate the source distribution accordingly. Crudely, this can be accomplished by introducing a weight $e^{-\eta^2/2\Delta^2}$ to any integral over source functions, which in the Hwa-Bjorken limit are uniformly distributed in $\eta$.

Modulations of the uniform Hwa-Bjorken source densities as described above neglect the effects of acceleration, as they continue to equate the coordinate $\eta$ with the collective rapidity. Since longitudinal acceleration requires a longitudinal pressure gradient, acceleration vanishes if the $\eta$ dependence is slow, or equivalently, if the reaction covers many units of rapidity. However, as the characteristic scale is between 1.5 and 2.0, acceleration cannot be easily dismissed.

\begin{figure}
\centerline{\includegraphics[width=0.5\textwidth]{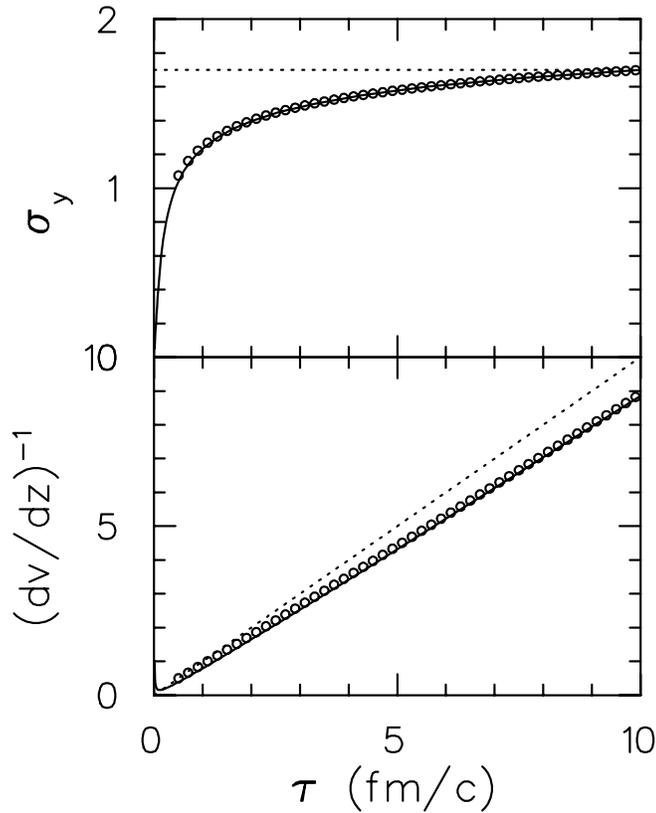}}
\caption{\label{fig:anal} In the upper panel, the rapidity width of the source distribution is shown for a numerical solution to one-dimensional hydrodynamics (circles) which assumed that collective rapidity was equal to $\eta$ at the initial time of $\tau_0=0.5$ fm/$c$. It closely matches an analytic solution (solid line) that assumed complete stopping but matched the velocity gradient of the numerical solution at $\tau_0$. This broadening is in stark contrast to the Bjorken-Hwa ansatz (dotted line) which assumes a fixed spread. For the same three cases, the inverse velocity gradient is shown in the lower panel. Whereas in the acceleration-less Hwa-Bjorken case the inverse velocity gradient equals $\tau$, it underestimates $\tau$ by of the order 10\% when acceleration is included.}
\end{figure}

Several groups are now performing hydrodynamic calculations in three dimensions \cite{nonaka,hirano,hama}. However, rather than using a full three-dimensional calculation to demonstrate the effects of longitudinal acceleration, I will employ simple one-dimensional solutions, where transverse degrees of freedom are neglected. In those solutions \cite{prattanal} one can compare the Hwa-Bjorken solution to one whose initial conditions, at some finite proper time $\tau_0$, were consistent with having the collective rapidity equal $\eta$. In the upper panel of Fig \ref{fig:anal}, one can see that the width of the rapidity distribution broadens substantially due to the acceleration. This has important implications for models of stopping, which in almost all cases assume the final rapidity distribution has not evolved from the stopping stage.

The inverse velocity gradient plays a critical role in two-particle correlation analyses, where the longitudinal size $R_{\rm long}$ is identified with the inverse velocity gradient,
\begin{equation}
R_{\rm long}=(dv/dz)^{-1} v_{\rm therm}.
\end{equation}
Here, $v_{\rm therm}$ is the longitudinal thermal velocity, which can be extracted from blast-wave analyses of spectra \cite{blastwave}. Given that $R_{\rm long}$ is directly taken from correlation analyses, one can extract the velocity gradient, which given the Hwa-Bjorken ansatz, $\tau=(dv/dz)^{-1}$, allows one to extract the breakup time. However, as can be seen in the lower panel of Fig. \ref{fig:anal}, this equivalence between $\tau$ and the inverse velocity gradient leads to an underestimate of the breakup lifetime on the order of 10\%. Since modelers always had a difficult time explaining how the sources grew to their large observed transverse sizes within what appeared to be 10 fm/$c$ or less, the extra 10\% is rather welcome. 

\section{Incorporating Viscosity}

Given the attention paid to viscosity in press releases announcing a ``perfect fluid'' at RHIC, it was surprising to see that none of the hydrodynamic treatments had incorporated viscous effects until the last year. Viscous shear increases the transverse pressure at the expense of the longitudinal pressure, thus giving transverse acceleration an early boost. This jump starts radial flow, and like longitudinal acceleration described above, helps explain correlation data. Although it is well understood that viscous shear can reduce elliptic flow, it is not clear that elliptic flow is reduced in all instances, and might even be increased by strong shear effects at very early times. 

It is worthwhile to first review the definitions and physics of viscosity. Rather than repeating the usual differential equations, I will write down the form of the stress energy tensor in the frame where the matter's velocity is small,
\begin{equation}
T^{\alpha\beta}=\left(\begin{array}{cccc}
\epsilon & (\epsilon+T_{xx})v_x & (\epsilon+T_{yy})v_y & (\epsilon+T_{zz})v_z\\
(\epsilon+T_{xx})v_x & T_{xx} & 0 & 0\\
(\epsilon+T_{yy})v_y & 0 & T_{yy} & 0\\
(\epsilon+T_{zz})v_z & 0 & 0 & T_{zz}
\end{array}\right)~~~~,
\end{equation}
where the coordinate axis have been chosen to diagonalize the stress energy tensor. It is always possible to diagonalize $T^{\alpha\beta}$ by boosting and rotating the coordinate system. The above form is easily derived by applying a small additional boost to the diagonalized form and keeping only those terms linear in the velocity. In this coordinate system the equations of motion $\partial_\alpha T^{\alpha\beta}$, 
\begin{eqnarray}
\label{eq:work}
(\partial_t\nabla\cdot v)\epsilon&=&-T_{xx}\partial_xv_x-T_{yy}\partial_yv_y-T_{zz}\partial_zT_{zz},\\
\label{eq:accel}
\partial_t v_i&=&-\frac{1}{\epsilon+T_{ii}}\partial_i T_{ii},
\end{eqnarray}
are physically transparent. Each element $T_{ii}$ plays the role of the pressure in the $i^{\rm th}$ direction. Equation (\ref{eq:work}) expresses that the work done in expanding a small element in the $i^{\rm th}$ direction is $T_{ii}dV$, and in Eq. (\ref{eq:accel}) the acceleration in the $i$ direction is governed by the gradient of $T_{ii}$.

Viscosity concerns the physics of how $T_{ij}$ differs from the equilibrium value $P(\epsilon)\delta_{ij}$. For Navier-Stokes, the difference is governed by the ansatz for the form of $T^{ij}$ in the frame of the matter,
\begin{equation}
\label{eq:navierstokes}
T_{ij}=P\delta_{ij}-\eta\left(\partial_iv_j+\partial_jv_i-(2/3)\nabla\cdot v\right)-B\nabla\cdot v,
\end{equation}
which should be true for sufficiently small velocity gradients. Here, the coefficients $\eta$ and $B$ are functions of the energy density and are referred to as the shear and bulk viscosities. Since the rate of change of the energy density is proportional to $\nabla\cdot v$, the bulk term describes how the effective pressure averaged over directions, $\bar{T}=(1/3)\sum T_{ii}$, departs from equilibrium due to a changing energy density. The strongest reason for $\bar{T}$ to depart from equilibrium is near $T_c$ where the chiral field, and perhaps the gluon condensate, are rapidly changing with $\epsilon$ \cite{kerstinvisc,kharzeevvisc}. Bulk viscosity will be discussed more heavily in the next section. The shear term describes the anisotropy in $T_{ij}$ due to the anisotropy of the velocity gradient, i.e., the traceless part of $T_{ij}$.

The Navier-Stokes equations suffer from a few shortcomings. First, numerical solutions include super-luminar modes \cite{muronga}. Secondly, forcing the stress energy tensor to be determined solely by the velocity gradient, without any influence from prior history or initial conditions, limits the range of physics that can be addressed. For these reasons, a class of equations referred to as Israel-Stewart equations have been considered and are being implemented by several groups \cite{israelstewart,heinz,romatschke,muronga}. In these approaches the elements $T_{ij}$ are treated as dynamical objects that exponentially relax toward Navier-Stokes values. In addition to the two new parameters describing the relaxation times, $\tau_\eta$ and $\tau_B$, solutions can also differ by varying the choice of how one scales the relaxing variable. For instance, one can define the variables as:
\begin{eqnarray}
a_{ij}&\equiv&\frac{T_{ij}-(1/3)\delta_{ij}\sum_iT_{ii}}{\tilde{h}_\eta(\epsilon)},\\
\nonumber
b&\equiv& \frac{\frac{1}{3}\sum_iT_{ii}-P}{\tilde{h}_B(\epsilon)},
\end{eqnarray}
which then decay towards their Navier-Stokes values,
\begin{equation}
\label{eq:israelstewart}
\dot{a}_{ij}=-\frac{(a_{ij}-a_{ij}^{N.S})}{\tau_\eta},~~~
\dot{b}=-\frac{(b-b^{N.S})}{\tau_B}.
\end{equation}

The usual choice for $\tilde{h}_\eta$ and $\tilde{h}_B$ is that they are set to unity. Another choice might be to set $\tilde{h}=P$. The choices can yield quite different behavior for large departures from equilibrium, which can occur for rapid expansions. For instance, if a component $T_{ii}$ is half of the equilibrium value when a rapid expansion lowered the energy density to the point where the pressure was half of the original value in much less than the relaxation time, choosing $\tilde{h}=1$ would lead to $T_{ii}$ falling to zero to maintain the offset with the equilibrium pressure. In contrast, if $\tilde{h}$ were chosen to be equal to the pressure, the same rapid expansion would maintain the ratio $T_{ii}/P=1/2$. For a departure from equilibrium due to a slowly relaxing mean field, the first choice $\tilde{h}_B=1$ might be physically justified since freezing the field freezes the pressure contribution to the field. In contrast, when one considers the shear contribution from particles that cannot maintain equilibrium due to a low scattering rate, it would be the ratio of the anisotropy of $T_{ij}$ to the kinetic pressure that should freeze out. In fact, the choice of $\tilde{h}_\eta=1$ could even result in negative components of $T_{ij}$ which should be unphysical for a dilute gas. This would suggest setting $\tilde{h}_\eta$ to the kinetic pressure, or perhaps to the enthalpy $h$.

In addition to the choice for $\tilde{h}$, one could also vary the target values for the offsets from the Navier-Stokes values assumed in Eq. (\ref{eq:israelstewart}). Rather than allowing the target values to become arbitrarily strong for large velocity gradients, as is the case for the Navier-Stokes equations, one could choose a target that saturates for large velocity gradients, e.g., $b^{\rm target}=b^{\rm sat.}\tanh(b^{\rm target}/b^{\rm sat})$. For instance, if the system behaves like a free gas, one might enforce the pressure to remain positive, or if the matter is dominated by classical color electric fields, one might require the components of $T_{ij}$ to be between $-\epsilon$ and $\epsilon$.

Since the components of $T_{ij}$ are dynamical objects in Israel-Stewart calculations, they can also accommodate arbitrary initial conditions. For instance, if the energy density is initially dominated by large electric fields, the stress energy tensor approaches a form, $T_{zz}=-\epsilon, T_{xx}=T_{yy}=\epsilon$. From this perspective, Israel-Stewart hydrodynamics is not so much a theory, but rather a surrogate model. Whereas for modest expansion rates Israel-Stewart theories reproduce Navier-Stokes hydrodynamics, for large expansion rates, the forms and parameters of the equations could be adjusted to mimic the behavior of a more sophisticated theory. For instance, if one is solving the equations of motion for the stress-energy tensor for Yang-Mills fields in a simple box geometry, one could tune the Israel-Stewart model to reproduce the evolution in the simple geometry.  The Israel-Stewart model could then be tuned to more complicated geometries. Since the equation $\partial_\alpha T^{\alpha\beta}=0$ is always true, and since varying the physical assumptions about the microscopic structure of the matter and equilibrium only affect the dynamics of $T_{ij}$, any form of Israel-Stewart equations that correctly mimics the evolution of $T_{ij}$, will correctly replicate the evolution of the entire stress energy tensor and of the collective velocity. 

The importance of how one chooses to formulate the Israel-Stewart approach and its parameters becomes clear when one considers the strength of the Navier-Stokes corrections at early times. Even for the proposed minimal-viscosity limit, $\eta=s/4\pi$, the Navier-Stokes correction to $T_{zz}$ becomes
\begin{equation}
T_{zz}^{N.S}=P-\frac{s}{4\pi}\frac{4}{3}\partial_zv_z\approx P\left(1-\frac{4}{3T \tau}\right),
\end{equation}
where the last step assumed an ideal gas where the entropy density is related to the pressure by $P=4sT$, and $\tau$ is the time. If the temperature is 300 MeV at $\tau=0.5$ fm/$c$, then $T_{zz}-P=-1.75P$, and the viscous correction is larger than the pressure itself. 

The principal manifestations of shear viscosity are at early times, where the large velocity gradients, $\partial_zv_z=1\tau, \partial_xv_x=\partial_yv_y=0$, is highly anisotropic. For free particles, the anisotropy at early times is bounded by the limit $T_{xx}=T_{yy}=\epsilon/2, T_{zz}=0$, whereas for longitudinal classical fields, the limit is $T_{xx}=T_{yy}=\epsilon, T_{zz}=-\epsilon$. For fields, the signs of the stress energy tensor can be understood by considering the expansion of the volume between two capacitor plates. Whereas pulling the plates apart requires work, expanding the area of the plates lowers the energy. Thus, during the first $\tau_0=0.5$ fm/$c$, the collective transverse velocities will expand by a characteristic amount
\begin{equation}
\bar{v}_\perp(\tau_0) \approx \frac{T_{xx}}{(\epsilon+T_{xx})R}\tau_0,
\end{equation}
where $R$ is a characteristic scale driving the transverse gradients. For off-center collisions $R$ could be $\sim$ 2 fm, $\bar{v}_\perp$ could be between 0.8333 (particles) and 0.125 (longitudinal fields). It is clear that the initial transverse impulse in the pre-equilibrated stage is non-negligible and could provide 10-15\% of the total collective flow generated during the expansion. The importance of this initial impulse in elliptic flow calculations remains to be analyzed in detail.

\section{Coupling to dynamic fields and bulk viscosity}

Viscous effects derive from a system's inability to maintain an equilibrium value, $P\delta_{ij}$, for the stress energy tensor at a fixed energy density, and a bulk viscosity ensues when $\sum_i T_{ii}/3\ne P$. For some cases, the bulk viscosity is manifestly zero. For instance, for a gas of free particles stress energy tensor becomes
\begin{equation}
T_{ij}=\sum_{\ell} \frac{p^{(\ell)}_ip^{(\ell)}_j}{E^{(\ell)}V},
\end{equation}
where $\ell$ refers to the individual particles within the volume $V$. For a gas of ultrarelativistic particles $\sum_{i} T_{ii}=\epsilon$, regardless of the momentum distribution and $\sum_i T_{ii}=\epsilon$. The same can be said for the non-relativistic case once the particle number is fixed, and $\epsilon$ refers to the kinetic energy density. Only a modest bulk viscosity will ensue if the distribution is semi-relativistic. Non-equilibrium chemistry can also lead to a bulk viscosity. For instance, in a cooling pion gas, the pion number would fall to maintain equilibrium as the temperature is near or below the pion mass \cite{gonggreinermuller,prattpiongas}. Therefore, a rapid expansion, or equivalently a large $\nabla\cdot v$, will result in over-population of the pion number. As the extra pions reduce the net kinetic energy available at a fixed total energy, this can be considered as a viscous effect. However, if a model dynamically accounts for chemistry by assigning currents for non-equilibrated particle numbers, the non-equilibrated chemistry no longer contributes to the bulk viscosity.

The most potentially important non-trivial sources of bulk viscosity derive from the vacuum condensates and fields which, near $T_c$, require a finite time to adjust to a rapidly changing energy density \cite{kerstinvisc}. For a field obeying a differential equation,
\begin{equation}
\partial_t^2\phi+\Gamma\partial_t\phi-m_{\rm eff}^2(\phi-\phi_{\rm equil})\phi=0,
\end{equation}
the drag term will pull the field away from the equilibrium value by an amount,
\begin{equation}
\phi-\phi_{\rm equil}=-\frac{\Gamma}{m_{\rm eff}^2}\dot{\phi_{\rm equil}}.
\end{equation}
This result can be understood by considering the analogous problem of a harmonic oscillator moving through a medium that provides a drag force. Given that the equilibrium value of the field changes at a rate determined by the rate of change of the entropy density $\dot{s}=-s\nabla\cdot v$,
\begin{equation}
\dot{\phi}_{\rm equil}=-\frac{d\phi_{\rm equil}}{d s}s\nabla\cdot v,
\end{equation}
one can identify the bulk viscosity as
\begin{equation}
\label{eq:Bfields}
B=\frac{d\phi_{\rm equil}}{ds}\frac{s\Gamma}{m_{\rm eff}^2}\left.\frac{\partial P}{\partial\phi_{\rm eq}}\right|_\epsilon.
\end{equation}
\begin{figure}
\centerline{\includegraphics[width=0.5\textwidth]{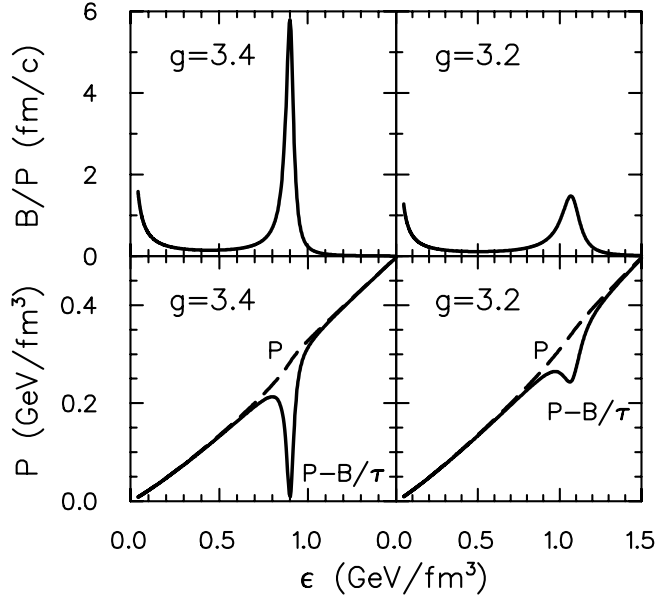}}
\caption{\label{fig:kerstin}The bulk viscosity due to non-equilibrium chiral fields are shown in the upper panels for two values of the parameter $g$ from Eq. (\ref{eq:chiral}). For $g=3.4$, the system is close to being a first-order transition and $\sigma$ changes rapidly as a function of energy density, and the bulk viscosity as calculated with Eq. (\ref{eq:Bfields}) has a sharp peak. Assuming a simple one-dimensional Bjorken expansion provides the conditions for calculating the Navier-Stokes expectation for $\bar{T}=P-B\nabla\cdot v$ in the lower panel. The very sudden changes in $\bar{T}$ invalidate the Navier-Stokes approach and suggest that either an Israel-Stewart approach, or a simultaneous solution of hydrodynamic and field equations of motion, should be undertaken.
}
\end{figure}

Since the effective mass will go to zero in the vicinity of a second-order phase transition, the viscosity will peak sharply at $T_c$ \cite{kerstinvisc}. Fig. \ref{fig:kerstin} shows the bulk viscosity as a function of energy density for a quark gas in the linear sigma model,
\begin{equation}
\label{eq:chiral}
H=-\frac{1}{2}\sigma\nabla^2\sigma+\frac{\lambda^4}{4}\left(
\sigma^2-f_\pi^2+m_\pi^2/\lambda^2\right)^2-h_q\sigma
+H_{\rm quarks}(m=g\sigma),
\end{equation}
After setting $\lambda=1/40$, the system has a first-order phase transition for $g>3.554$, and a cross-over for lower $g$. The bulk viscosity, scaled by the pressure, is displayed in the upper panels of Fig. \ref{fig:kerstin} for $g=3.4$ and $g=3.2$. As expected, the peak in the bulk viscosity is stronger for $g$ near the critical value as that corresponds to a more rapid change in the $\sigma$ condensate as a function of energy density. In the lower panel, the effect on $\bar{T}$ is shown for the two values of $g$ in a simple one-dimensional Hwa-Bjorken example. It is clear that viscous effects can be large, and for the $g=3.4$ case, so large that Navier-Stokes approaches are questionable. Solving Israel-Stewart equations should effectively smooth out the bumps in $\bar{T}$, but as emphasized above, the form of the equations and the parameters should be carefully matched to reproduce the expected dynamics for the fields. 

Rather than solving Israel-Stewart equations, another option is to dynamically solve the equations of motion of the chiral field. This was done in \cite{paechdumitru}, though the equations ignored the drag term which from the considerations above should be important. The one crucial requirement for Israel-Stewart approaches to be valid is that equilibrium should be approached exponentially. In applying equations of motion for the field, the behavior is more of an oscillatory nature, although in the limit of strong damping the behavior becomes exponential. Thus, the appropriateness of Israel-Stewart approaches for modeling non-equilibrium behavior near $T_c$ is not yet clear. However, explicity modeling the dynamics of the fields, alongside solving the hydrodynamic equations of motion, should always be tenable.

We conclude with speculation regarding the effects of the non-equilibrium behavior near $T_c$. As can be seen from Fig. \ref{fig:kerstin}, the effect should be similar to what one would get from softening the equation of state near $T_c$, and perhaps even creating an unstable region with an effectively negative speed of sound. Lowering the speed of sound at the periphery of the collision volume has the effect of building up a density inversion, and thus a sudden dissolution of the system. This behavior should push the resulting $R_{\rm out}/R_{side}$ ratio from two-particle correlations downward, and perhaps in line with the low value observed experimentally.

\section*{Acknowledgments}
Support was provided by the United States Department of Energy, Grant No. DE-FG02-03ER41259.

\end{document}